\documentclass[superscriptaddress,twocolumn,10pt,pra,aps,showpacs,longbibliography,floatfix]{revtex4-2}

\usepackage[dvipsnames]{xcolor}

\definecolor{darkred}{rgb}{0.6,0.05,0.05}
\definecolor{darkgreen}{rgb}{0.05,0.6,0.05}
\definecolor{darkblue}{rgb}{0.05,0.05,0.6}
\usepackage[colorlinks]{hyperref}
\hypersetup{colorlinks,
linkcolor=darkred,
filecolor=darkgreen,
urlcolor=darkblue,
citecolor=darkgreen}

\usepackage{mathtools}
\usepackage{amsmath}
\allowdisplaybreaks
\usepackage{ bbold }
\usepackage{amssymb}
\usepackage[normalem]{ulem}
\usepackage{amsthm}
\usepackage{xfrac}
\usepackage{dsfont}
\usepackage{siunitx}
\usepackage{physics}
\usepackage{graphicx}% Include figure files
\usepackage{hyperref}% add hypertext capabilities
\usepackage[capitalise]{cleveref}
\usepackage{dcolumn}% Align table columns on decimal point
\usepackage{bm}% bold math
\usepackage{layouts}
\usepackage{comment}

\renewcommand{\paragraph}[1]{\noindent \textbf{#1}---.}

\begin{document}

\title{Chaos in Time: Incommensurable Frequencies and Dissipative Continuous Quasi Time Crystals}

\title{Chaos in Time: A Dissipative Continuous Quasi Time Crystals}

\author{Parvinder Solanki}
\email{parvinder@mnf.uni-tuebingen.de}
\affiliation{Institut f\"ur Theoretische Physik, Universit\"at T\"ubingen,
Auf der Morgenstelle 14, 72076 T\"ubingen, Germany}
\thanks{Part of this work was done at {Department of Physics, University of Basel, Klingelbergstrasse 82, CH-4056 Basel, Switzerland}.}

\author{Fabrizio Minganti}
\email{fabrizio.minganti@gmail.com}
\affiliation{Institute of Physics, \'{E}cole Polytechnique F\'{e}d\'{e}rale de Lausanne (EPFL), 1015 Lausanne, Switzerland}
\thanks{Currently at {Alice \& Bob,  53 Boulevard du G\'en\'eral Martial Valin, 75015, Paris, France}.}

\begin{abstract}
While a generic open quantum system decays to its steady state, continuous time crystals (CTCs) develop spontaneous oscillation and never converge to a stationary state.
Just as crystals develop correlations in space, CTCs do so in time.
Here, we introduce a \textit{Continuous Quasi Time Crystals} (CQTC). 
Despite being characterized by the presence of non-decaying oscillations, this phase does not retain its long-range order, making it the time analogous of quasi-crystal structures.
We investigate the emergence of this phase in a system made of two coupled collective spin sub-systems, each developing a CTC phase upon the action of a strong enough drive.
The addition of a coupling enables the emergence of different synchronized phases, where both sub-systems oscillate at the same frequency.
In the transition between different CTC orders, the system develops chaotic dynamics with aperiodic oscillations.
These chaotic features differ from those of closed quantum systems, as the dynamics is not characterized by a unitary evolution.
At the same time, the presence of non-decaying oscillations makes this phenomenon distinct from other form of chaos in open quantum system, where the system decays instead.
We investigate the connection between chaos and this quasi-crystalline phase using mean-field techniques, and we confirm these results including quantum fluctuations at the lowest order.
\end{abstract}

\maketitle

\paragraph{Introduction}
Classical nonlinear systems produce a variety of phenomena, including instabilities, period multiplication,  and synchronization \cite{strogatz2018nonlinear}. 
Similar phenomena also occur in quantum mechanical systems in the presence of dissipation---the non-unitary dynamics 
ensuing from the interaction between a quantum system and its environment \cite{Giorgi2019}.
A primary example of dissipation-induced phenomena are continuous time crystals (CTCs) \cite{iemini2018boundary, tucker2018shattered,buca2019non,Booker_2020,PhysRevLett.123.260401,zhu2019dicke,lledo2019driven,seibold2020dissipative,minganti2020correspondencedissipativephasetransitions,prazeres2021boundary,piccitto2021symmetries,carollo2022exact,krishna2022measurement,seeding2022michal,hurtado2020raretimecrystal,PhysRevB.108.024302,PhysRevLett.127.133601,iemini2024sectors,cabot2023nonequilibrium,mattes2023entangled,solanki2024exotic,Mukherjee2024correlations}, states that like ``standard'' crystals develop long-range order, but realizes it in time rather than in space.
Normally, an open quantum system evolves toward its unique steady state and, upon reaching it, the dynamics become stationary. CTCs breaks this time-translational invariance, maintaining an oscillatory behavior even at infinite times.
This space-to-time analogy extends to several interacting CTCs, as oscillating phases combine or melt, losing all-time periodicity or synchronizing with novel ones \cite{seeding2022michal,solanki2022role,solanki2024exotic,liu2024microwaveseedingtimecrystal}.
CTCs have been associated with enhanced sensing \cite{cabot2023continuous,montenegro2023quantum}, quantum-to-classical transition \cite{dutta2024quantumoriginlimitcycles} and non-equilibrium thermodynamic \cite{igor_thermodynamics,paulino2024thermodynamicscoupledtimecrystals,PhysRevA.108.023516},
prompting their experimental investigations \cite{kessler2021observation,kongkhambut2022observation,greilich2024robust}.

\begin{figure}[htp!]
    \centering
    \includegraphics[width=0.99\linewidth]{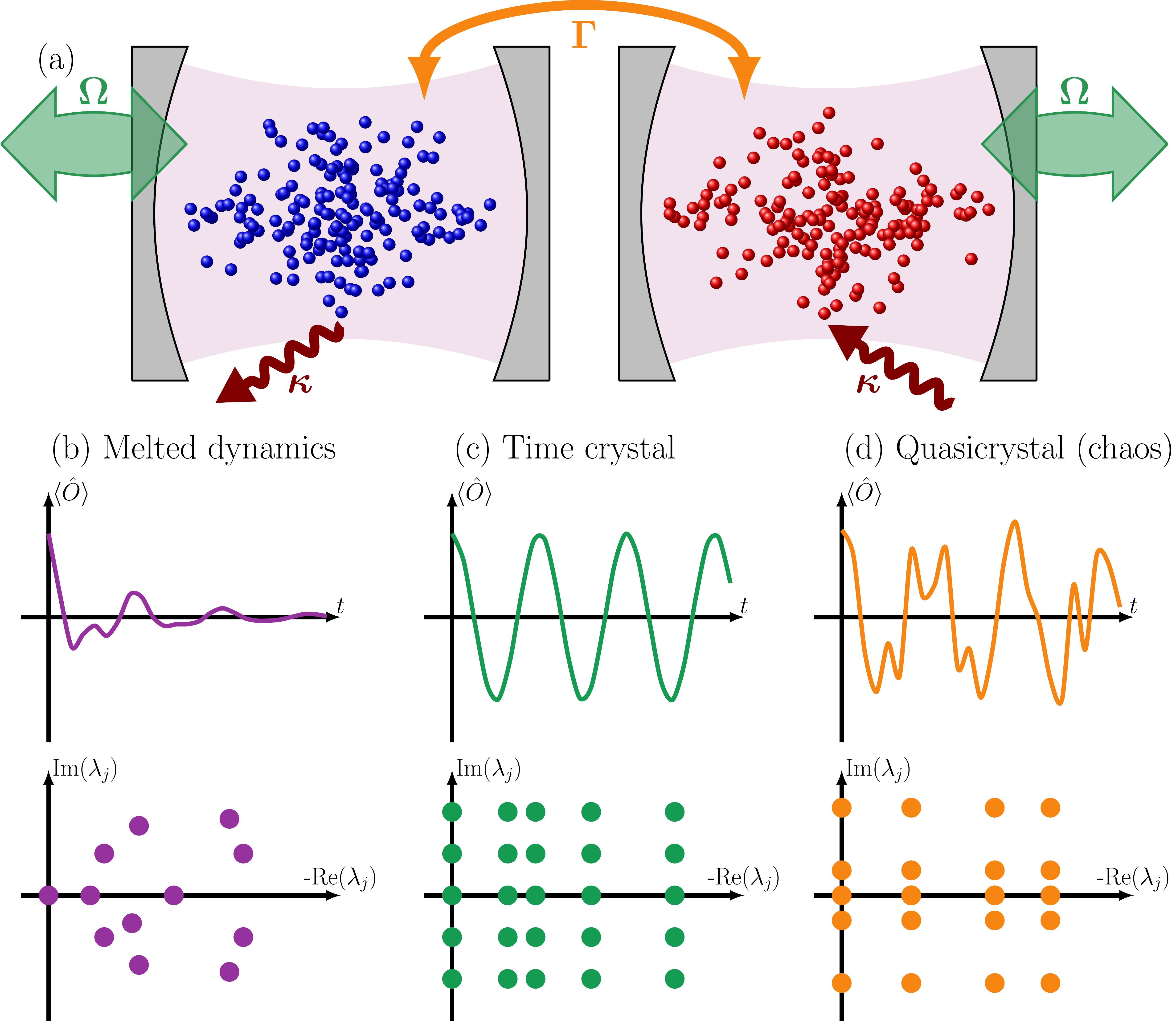}
    \caption{
    Sketch of the model and emergent phases. (a) Two systems, each described by the Dicke model in Eq.~\eqref{Eq:Driven_Dissipative_Dicke}, are coupled and result in Eq.~\eqref{eq:ME_spins}. According to the choice of parameters and the relative strength of the coupling, the collective phase of the system changes between 
    (b) Melted: Irreversible decay towards the steady state. 
    When analyzing the structure of the Liouvillian superoperator [c.f. \eqref{Eq:spectrum}], all eigenvalues are characterized by Re$(\lambda_j)<0$, except the one associated with the steady state Re$(\lambda_j)=0$. 
    (c) Time Crystal: Undamped, regular oscillations. This results in Re$(\lambda_j)=0$ and equally spaced Im$(\lambda_j)\neq0$ for a portion of eigenvalues. (d) Quasicrystal: Persistent but aperiodic oscillations. Here, Re$(\lambda_j)=0$ and non-commensurate Im$(\lambda_j)$.}
    \label{fig:sketch}
\end{figure}

Chaos--a prerogative of nonlinear systems--is an amplification of the incertitude on the system's initial state, making it so that ``the present determines the future but the approximate present does not approximately determine the future'' \cite{lorenz1963deterministic}.
Despite having a linear dynamics, quantum systems can display a certain degree of randomness, leading to parallels between classical and quantum chaos. 
The features and characteristics of these quantum chaotic systems have been characterized via random matrix theory \cite{jensen1992quantum,lakshminarayan2002chaos, bohigas_characterization_1984, haake_quantum_2001, dalessio_quantum_2016}.
These definitions have been extended to open quantum systems \cite{akemann2019universal,lucas2020complex,kawabata2023symmetry,li2024spectralformfactorchaotic}, with recent investigation focusing on the stochastic nature of the system-environment interaction \cite{ferrari2023steady,dahan_classical_2022,ferrari2024chaosspatialprethermalizationdrivendissipative}.

A distinctive feature of chaos is the presence of strange attractors in phase space, points around which a system describes aperiodic orbits \cite{strogatz2018nonlinear}.
A key question is whether this chaotic feature can emerge in dissipative configurations, with a system not relaxing to its steady state but maintaining aperiodic oscillations. 
Given the correspondence between space and time crystals, this phase would represent an ordered but non-periodic structure in time, thereby creating a dissipative \textit{continuous quasi-time crystal} (CQTC).

We present evidence for the existence of CQTCs and characterize them. 
We analyze this phase using mean-field techniques, and we verify that beyond-mean-field techniques reproduce the mean-field results in the thermodynamic limit.
We thus show that the resulting phase is robust with respect to finite-size effects and analyze how the emission spectrum could detect the appearance of this chaotic behavior.
As this phase leads to chaotic oscillatory behavior within a CTC, this problem can be seen as chaos in a 1D portion of an otherwise 2D spectrum, making its features distinct from both the prediction of Hamiltonian or Liouvillian chaos.

\bigskip

\paragraph{Model} As sketched in Fig. \ref{fig:sketch}, the system we consider is two coupled driven-dissipative Dicke models. 
Each one of them is made of $N$ all-to-all connected spins coupled to distinct baths, subjected to an external drive of amplitude $\Omega$ and collective decay of rate $\kappa$ \cite{iemini2018boundary}.
If the coupling $\Gamma=0$, a single Dicke model is described by the master equation
\begin{equation}
\label{Eq:Driven_Dissipative_Dicke}
    \dot{\rho}=-i[\Omega S_x,\rho]+\frac{\kappa}{S} D[S_\pm] \rho.
\end{equation}
Here, $S_{\eta}=\sum_i \sigma_\eta^i$ are collective spin operators, $S=N/2$ is the total spin size, and $\sigma_{\eta}^i$ are Pauli spin operators of the $i$th spin and $2 \sigma_\eta^{\pm} = \sigma_\eta^x \pm i \sigma_\eta^y$, with $\eta
\in \{x,y,z,\pm\}$. 
This system transits from a melted phase  [uncorrelated in time, Fig.~\ref{fig:sketch} (b)] for $\Omega/\kappa<1$, to the a CTC one [time correlated, Fig.~\ref{fig:sketch} (c)] for $\Omega/\kappa>1$, also known as a boundary time crystal. 
Although this transition occurs only in the thermodynamic limit $S\rightarrow \infty$, increasing the system's size $N$ results in longer-lived oscillations, signaling the development of long-range order in time.

When $\Gamma\neq 0$ and the two sub-systems are coupled, they evolve as
\begin{align}
    \dot{\rho}=-i[ \Omega (S_x^A &+ S_x^B) +\frac{\Gamma}{S}(S_+^A S_-^B+S_-^AS_+^B),\rho] \nonumber \\  &+ \frac{\kappa}{S} D[S_-^A]\rho + \frac{\kappa}{S} D[S_+^B]\rho.\label{eq:ME_spins}
\end{align}
Compared to two copies of the model in Eq.~\eqref{Eq:Driven_Dissipative_Dicke}, aside from the coupling $\Gamma$, we assume that the $B$ model experiences gain rather than dissipation.

\bigskip

\paragraph{Phase diagram} 
 We construct a phase diagram by analyzing the mean-field dynamics of the system (we refine and verify this analysis below).
The mean-field approximation consists in neglecting correlations between the spin operators, so that $ \langle S_\nu^i  S_\mu^j \rangle = \langle S_\nu^i  \rangle  \langle S_\mu^j \rangle$.
This method is expected to be valid in the thermodynamic limit $S\rightarrow \infty$ \cite{HuybrechtsPRB20, CarolloPRL21}, 
and has been applied both in the study of CTCs \cite{iemini2018boundary,carollo2022exact,seeding2022michal,solanki2024exotic} and in that of chaos \cite{prazeres2021boundary,li2022nonlinear,solanki2024exotic}. 
The mean-field dynamics of the re-scaled spin operators $m_\nu^A = \langle S_\nu^A \rangle/S$ then reads
\begin{align}
    \frac{dm_x^{A} }{dt} & = \kappa_A m_x^{A}  m_z^{A}  + \Gamma m_z^{A} m_y^{B} , \nonumber\\
    \frac{dm_y^{A} }{dt} & = -\Omega m_z^{A}  + \kappa_A m_y^{A}  m_z^{A}  - \Gamma m_z^{A}   m_x^{B} , \nonumber\\
    \frac{dm_z^{A} }{dt} & = \Omega m_y^{A}  - \kappa_A [ (m_x^{A})^{2} + (m_y^{A})^2 ]\nonumber \\
    & \qquad+ \Gamma (m_y^{A}   m_x^{B}  - m_x^{A}  m_y^{B} ), \label{eq:mf3}
\end{align}
where $\kappa_A=\kappa$. The same equation holds for the $B$ mode, swapping the indices $A$ and $B$, and $\kappa_B=-\kappa$ (see Appendix \ref{mean-field}).

\begin{figure}[t!]
    \centering
    \includegraphics[width=\linewidth]{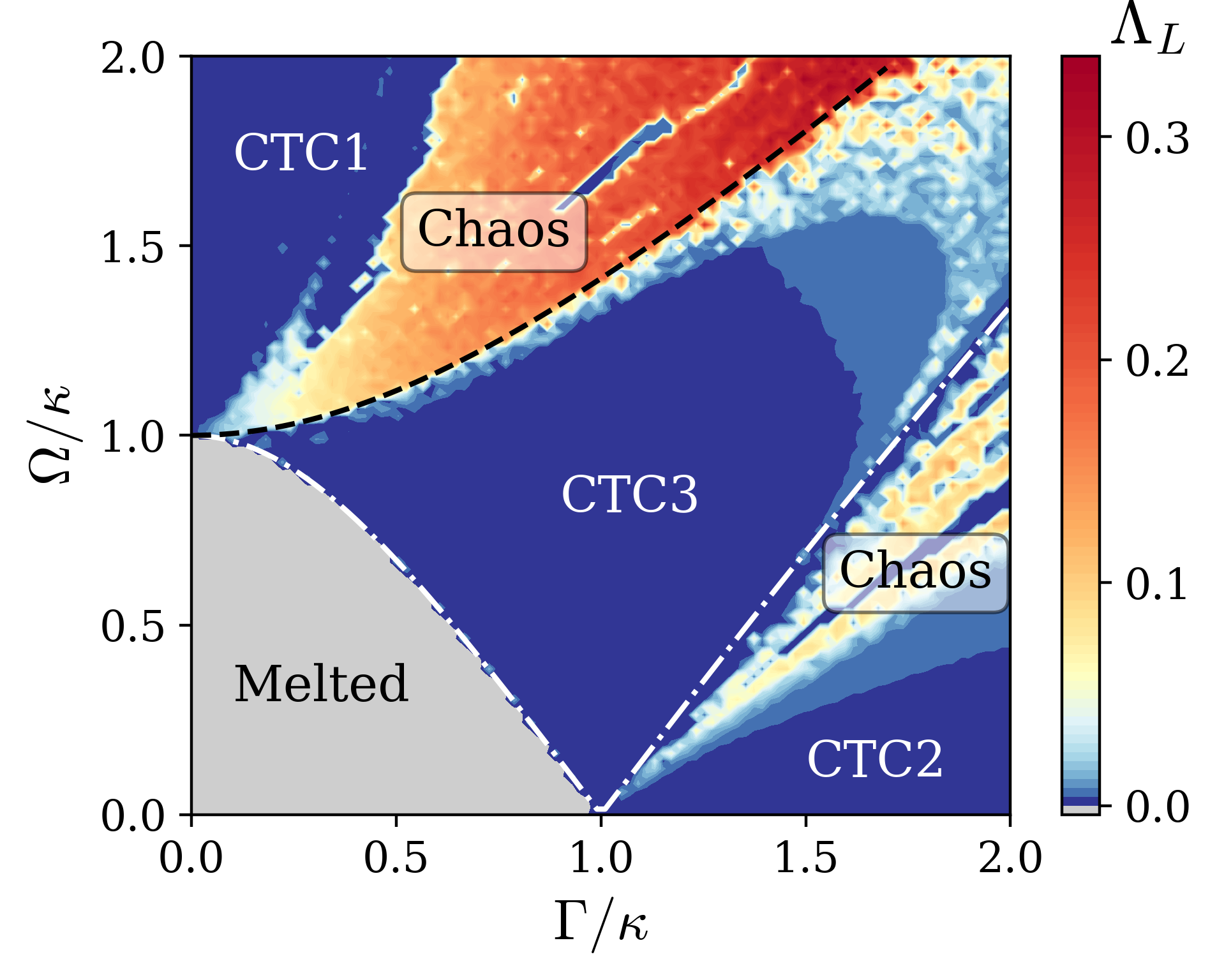}
    \caption{\textbf{Phase diagram}: Characterization of different phases using the largest Lyapunov exponent $\Lambda_L$. Chaotic dynamics, characterized by $\Lambda_L>0$, emerge at the interface of different time crystal phases CTC1, CTC2, and CTC3, described by $\Lambda_L=0$.
    The melted phase is given by $\Lambda_L<0$. The dashed black and the white dotted-dashed lines separate the CTC from the chaotic CQTC phases, and have been obtained by fixed point analysis.}
    \label{fig:phase}
\end{figure}
\begin{figure*}[htp!]
    \centering
    \includegraphics[width=\linewidth]{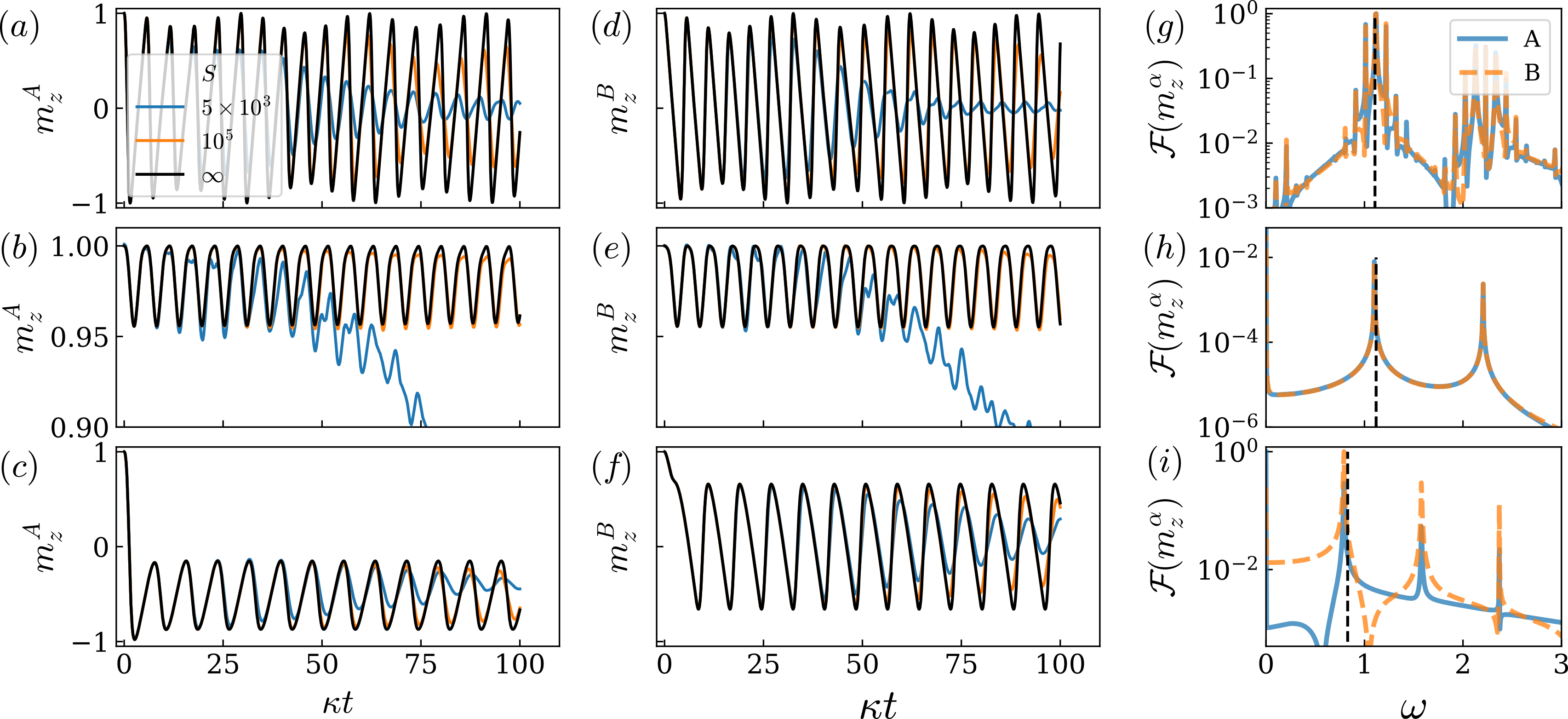}
    \caption{\textbf{Time crystal phases}: Time evolution and the emission spectrum of different CTC phases using the mean-field and the truncated Wigner approximation (TWA). Panels (a,d), (b,e) and (c,f) represents the time evolution of $m_z^{A,B}$ of CTC1 ($\Omega/\kappa=1.5,\Gamma/\kappa=0.1$), CTC2 ($\Omega/\kappa=0.1,\Gamma/\kappa=1.5$) and CTC3 ($\Omega/\kappa=0.75,\Gamma/\kappa=0.75$), respectively, for different system sizes. Transient behavior can be observed for finite system sizes. The normalized emission spectra of the corresponding mean-field dynamics is given by (g-i). 
    }
    \label{fig:TC_phase}
\end{figure*}
To determine the phases of the system, we compute the largest Lyapunov exponent $\Lambda_L$  (see Appendix \ref{lyapunov}).
$\Lambda_L$ measures the sensitivity of the dynamics of a nonlinear system to its initial condition.
Two nearby trajectories diverge if $\Lambda_L>0$, and the system becomes chaotic.
Limit cycles and the CTCs correspond to $\Lambda_L=0$. 
Finally, if $\Lambda_L<0$, the system is in the melted phase, converging to a fixed point.
The phase diagram is shown in Fig.~\ref{fig:phase}.
To further characterize the CTC phases, we also resort to fixed point analysis.
Namely, we find the fixed points $\dot{m}^{A,B}_{x,y,z}=0$, and by computing the corresponding Jacobian eigenvalues, we characterize the flow around them (see Appendix \ref{mean-field} for more details).
Depending on the choice of parameters, our model exhibits various dynamical phases, namely (see Appendix \ref{mean-field}):
\vspace*{5pt}

\noindent \textit{Melted phase:} For $\Omega< \vert \Gamma^2-\kappa^2 \vert/\sqrt{\Gamma^2+\kappa^2}$ and $\Gamma/\kappa <1$, $\Lambda_L<0$ and the system exhibits a time invariant steady with $m_z^{A,B}\neq 0$. The time evolution is sketched in Fig.~\ref{fig:sketch}(b). 
\vspace*{5pt}

\noindent \textit{Continuous time crystal phases:} Here, $\Lambda_L=0$ indicates oscillations without amplification or damping [c.f. the sketch in Fig.~\ref{fig:sketch}(c)]. We identify three different types of CTCs.
\vspace{3pt}
\\
\textbullet  (CTC1): Standard time crystal.  If $\Gamma =  0$, CTCs emerge in each system as a result of competition between the external drive strength $\Omega$ and the dissipation rate $\kappa$  as for the model in Eq.~\eqref{Eq:Driven_Dissipative_Dicke}. Also, the fixed point is similar to that of the boundary time crystal introduced in Ref.~\cite{iemini2018boundary}, and thus both sub-systems oscillate around $m_z^{A,B}=0$.
CTC1 is robust to the perturbation by some small coupling value $\Gamma/\kappa \neq 0$, even if the dynamics becomes quasi-periodic with more than one frequency, as shown in Fig.~\ref{fig:TC_phase}(a, d). 
\vspace{3pt}
\\
\textbullet  (CTC2): Phase-locked time crystal. Also in the opposite limit of $\Omega = 0$ the system can transition from a melted ($\Gamma/\kappa<1$) to a CTC phase ($\Gamma/\kappa>1$). 
In this case, the time crystal phase emerges as a consequence of coherent interaction between two subsystem $A$ and $B$, which experience loss and gain, respectively. 
This phase is more similar to a lasing phase, and the emergent CTC resemble to that observed in $U(1)$ symmetry breaking \cite{minganti2020correspondencedissipativephasetransitions}.
This is a new time crystal phase, where both sub-systems synchronously oscillate around non-zero $m_z^{A,B}$ values. 
Also, in this case, the phase is robust to perturbation, and the CTC3 phase is present for $\Omega< \vert \Gamma^2-\kappa^2 \vert/\sqrt{\Gamma^2+\kappa^2}$ and $\Gamma/\kappa >1$.
\vspace{3pt}
\\
\textbullet  (CTC3): Asymmetric time crystal. If none of the two mechanisms dominate over the other and $\Gamma \simeq \Omega$, a new asymmetric time crystal phase emerges, where sub-system $A$ oscillates around non-zero $m_z^A$ value, and the sub-system $B$ around $m_z^B=0$.  
The asymmetry is also evident in the counter-phase oscillation of the two crystals.
The boundary of this phase lie  at $\vert \Gamma^2-\kappa^2 \vert/\sqrt{\Gamma^2+\kappa^2}\leq \Omega \leq \sqrt{\Gamma^2+\kappa^2}$.
\vspace*{5pt}

\begin{figure}[t!]
    \centering
    \includegraphics[width=\linewidth]{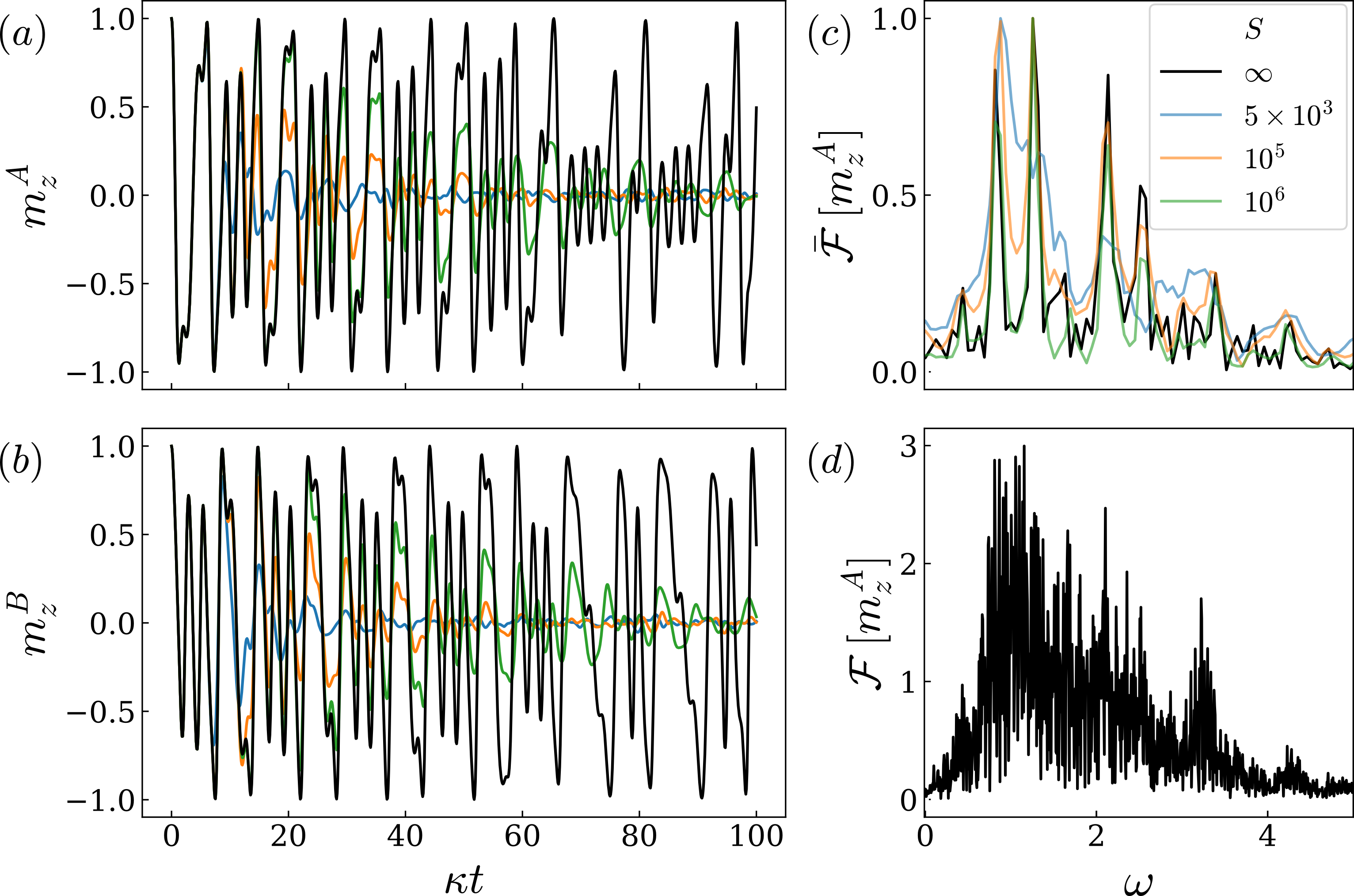}
    \caption{\textbf{Chaos}: Panels (a) and (b) illustrate the time evolution of subsystems $A$ and $B$ using mean-field dynamics and TWA, with parameters $\Omega/\kappa=2$ and $\Gamma/\kappa=1.4$. (c) and (d) The normalized emission spectrum for the dynamics of $m_z^{A}$ shown in (a). The data in (c) have been obtained by the Fourier transform in \eqref{Eq:Observable_time} over a time $\kappa t = 10^2$, while those in (d) for a time $\kappa t = 10^3$. 
    }
    \label{fig:chaos_features}
\end{figure}

\noindent \textit{Chaotic quasi-crystal phases:}  
The system exhibits chaotic dynamics at the interfaces of two time crystal phases, as signaled by $\Lambda_L>0$. 
The system does not decay to a single fixed point and maintains oscillations, but these are not periodic, as depicted in Fig.~\ref{fig:sketch}(c).
The competition between the different ordering mechanisms induced by $\Omega$ and $\Gamma$ is the cause of the chaotic dynamics.
For instance, consider $\Gamma/\kappa>1$ and $\Omega/\kappa<1$.
By considering the Eq.~\eqref{eq:mf3}, a tension in the system emerges;  while $\Gamma$ imposes oscillations where both $m_x$ and $m_y$ participate, $\Omega$ alone would impose an oscillation along the $y-z$ plane.
Thus, when $\Omega$ becomes sufficiently large, the two mechanisms trying to impose different and non-commensurable orders to the system
lead to chaos in the CTC.
One such example is shown in Fig.~\ref{fig:chaos_features}.

\bigskip

\paragraph{Truncated Wigner Approximation} 
To verify the phase diagram, we go beyond the mean-field analysis and introduce truncated Wigner approximation (TWA) for open quantum spins \cite{Huber2021}.
First, we use the Schwinger transformation, defined for the $A$ spin as 
\begin{equation}
    S_+^{A}= a_1^\dagger a_2, ~~~S_-^{A}=a_1 a_2^\dagger, ~~~S_z^{A}=(a_1^\dagger a_1-a_2^\dagger a_2)/2,
\end{equation}
where $a_{1,\, 2}$ are annihilation operator. The same relation holds true for the $B$ system and the corresponding $b_{1,\, 2}$ operators.
Equation~(\ref{eq:ME_spins}) then becomes
\begin{align}
    \dot\rho&=-i[\frac{\Omega}{2} (a_1^\dagger a_2+a_1 a_2^\dagger+ b_1^\dagger b_2+b_1 b_2^\dagger)+\frac{\Gamma}{S}(a_1^\dagger a_2 b_1 b_2^\dagger\nonumber \\
    &+a_1 a_2^\dagger b_1^\dagger b_2),\rho]+\frac{\kappa}{S}(D[a_1 a_2^\dagger]\rho+D[b_1^\dagger b_2]\rho).
\end{align}

The TWA is then a technique to transform the time evolution of the bosonic system into a set of stochastic (It\^o) differential equations that, in our case, read
\begin{align}
    d \alpha_i &= -i\Big[\frac{\Omega}{2}+\frac{\Gamma}{S} (\delta_{i,1}\beta_1 \beta_2^*+\delta_{i,2}\beta_1^* \beta_2 ) \Big] \alpha_{j\neq i} dt + (-1)^{i} \frac{\kappa}{S} \nonumber \\&\Big(\vert \alpha_{j\neq i} \vert^2+\frac{1}{2}\Big)\alpha_i dt +\sqrt{\frac{\kappa}{2S}\Big(\vert \alpha_{j\neq i} \vert^2+\frac{1}{2}\Big)}d\mathcal{W}_{i},\\
    d \beta_i &= -i\Big[\frac{\Omega}{2}+\frac{\Gamma}{S} (\delta_{i,1}\alpha_1 \alpha_2^*+\delta_{i,2}\alpha_1^* \alpha_2 )\Big]\beta_{j\neq i} dt - (-1)^{i} \frac{\kappa}{S}\nonumber \\&\Big(\vert \beta_{j\neq i} \vert^2+\frac{1}{2}\Big)\beta_i dt +\sqrt{\frac{\kappa}{2S}\Big(\vert \beta_{j\neq i} \vert^2+\frac{1}{2}\Big)}d\mathcal{W}_{i}.
\end{align}
Here $\delta_{i,j}$ is a delta function and $d\mathcal{W}_{i}=d\mathcal{W}_{i1}+id\mathcal{W}_{i2}$ are complex independent Wiener processes with $\langle d\mathcal{W}_{i1} d\mathcal{W}_{i2}\rangle=\delta_{12}dt$ (see Appendix \ref{TWA-appendix} for the detailed derivation).
Correspondence between the expectation values of the system's observable and the outcome of the stochastic trajectories can be established, as $\bar{\alpha_i} = \langle a_i \rangle$, $\bar{\beta_i} = \langle b_i \rangle$, with $\bar{\alpha_i}$ indicating the average over multiple noise configurations.
While the exponential complexity of the quantum systems gets reduced to a few coupled differential equations for complex numbers, the price to pay is that the TWA includes quantum fluctuations only at the lowest order in $\hbar$. 
Approaches based on TWA are expected to predict correlations beyond mean-field ones in regimes where nonlinearities are small, as shown in studies on CTCs, solitons, and phase transitions in a dissipative framework \cite{Foss-FeigPRA17,VicentiniPRA18,SeiboldPRA22}.

In Figs.~\ref{fig:TC_phase}~and~\ref{fig:chaos_features}, we confirm that all CTC and chaotic phases approach the mean-field limit upon increasing the system size.
We thus argue that they are stable even when including beyond mean-field terms.
In conclusion, the TWA analysis confirms the predictions of the phase diagram shown in Fig.~\ref{fig:phase}.

\bigskip

\paragraph{Emission spectra} Signatures of chaos or regular dynamics are captured by the Liouvillian spectrum, defined through
\begin{equation} \label{Eq:spectrum}
	\dot \rho = \mathcal{L} \rho, \, \, \rho(t) = e^{\mathcal{L} t} \rho(0), \, \,  \mathcal{L} \rho_j = \lambda_j \rho_j,
\end{equation}
where $\lambda_j$ are the eigenvalues, encoding the oscillation and decay frequencies, and $\rho_j$ the eigenoperators of the system describing the states explored.
Time crystalization can be seen as the presence of several purely imaginary eigenvalues, indicating oscillations without damping; as the oscillations are regular, all these eigenvalues must be equidistant, allowing only for a periodic motion to occur, as shown in Fig.~\ref{fig:sketch}(b). This is in contrast to the melted phase, where all eigenvalues except the one associated with the steady-state have a negative real part, indicating a decay [Fig.~\ref{fig:sketch}(c)].
As QCTC must still display no decay, it should also be characterized by eigenvalues with zero real parts. At the same time, these eigenvalues should not be periodically spaced, dictating the presence of incommensurate oscillations [see Fig.~\ref{fig:sketch}(d)].

Diagonalizing the Liouvillian is a daunting task, even for small system sizes.
However, any observable $\hat{O}$ must obey
\begin{equation}\label{Eq:Observable_time}
\langle \hat{O}(t)\rangle  = \sum_j c_j e^{\lambda_j (t)},
\end{equation}
where $c_j$ are coefficients depending on the observables, the left Liouvillian eigenoperators, and the initial state, while $\lambda_j$ are the Liouvillian eigenvalues in Eq.~\eqref{Eq:spectrum}. Fourier transforming this correlation function must result in a series of peaks centered around the frequency of oscillation of these modes, whose linewidth depends on the real part of the eigenvalue and on the total time the signal has been acquired. Thus, the presence of sharp peaks in the Fourier transform is a signature of any non-decaying oscillation.

We investigate the emission spectrum of the CTC phase, as shown in Fig.~\ref{fig:TC_phase}(g-i).
We observe that the system exhibits well-defined peaks, indicating that the phase is indeed coherent.
The frequency of oscillations matches the mean-field dynamics.
The main peak of the TC1 spectrum matches the eigenvalue of the Jacobian matrix $\pm i\sqrt{\Omega^2-\kappa^2}$ obtained by the fixed point analysis.
Similarly the frequency of phases TC2 and TC3 is given by $\pm i\vert\Gamma^2-\kappa^2\vert \sqrt{\frac{\Omega^2 (\Gamma^2+\kappa^2)}{(\Gamma^2-\kappa^2)^2}-1}$ and $\pm i\sqrt{\Gamma^2-\kappa^2}\sqrt{1-\frac{\Omega^2 (\Gamma^2+\kappa^2)}{(\Gamma^2-\kappa^2)^2}}$ respectively (see Appendix \ref{mean-field}).
This spectral gap closing with the equally spaced purely imaginary eigenvalues, matching the mean-field frequencies, have been discussed in previous studies \cite{iemini2018boundary,seeding2022michal,solanki2024exotic,dutta2024quantumoriginlimitcycles}.

We then compute the same quantity in the chaotic phase.
Given the unequally spaced feature of the spectrum, this chaotic behavior manifests itself as a broad random distribution in the emission spectra, as shown in Figs.~\ref{fig:chaos_features}(c-d).
Notice also that the chaotic behavior only becomes evident for large system sizes. 
Indeed, in finite-size simulation, the finite lifetime of oscillation makes it impossible to distinguish the details of the chaotic dynamics, as each peak in the emission spectrum is broadened by these finite-size effects.
We also stress the difference in Figs.~\ref{fig:chaos_features}(c-d), for which the signal acquisition time is different. The latter, having been acquired for longer time, displays significantly more structure, and shows the presence of many sharp peaks.

\bigskip

\paragraph{Conclusions}
In this work, we demonstrate that the interaction between two coupled time crystals reveals an intricate phase diagram, featuring three distinct time-crystalline orders, alongside a melted phase and a novel quasi-periodic (chaotic) structure. 
The key feature of our model lies in the presence of two competing mechanisms that induce different time orders within the system. The interplay between these mechanisms leads to a breaking of time-translational symmetry without establishing long-range order, resulting in persistent quasi-periodic or chaotic dynamics in time.

We tie the presence of CTCs and CQTC to the Liouvillian eigenspectrum, and provide numerical evidence for the signature of these phenomena in the emission spectra of these systems.
The eigenspectrum associated with the CQTC phase involves only a portion of the eigenvalues, a 1D sub-manifold in the larger 2D space of Liouvillian eigenvalues. This feature sets CQTC chaos apart from both Hamiltonian and Liouvillian chaos, where the entire spectrum contributes to chaotic dynamics, whether in the bulk of the Liouvillian superoperator or the Hamiltonian system. 
Characterizing randomness and complex behavior through random matrix theory in the portion of the spectrum associated with undamped oscillations is one of our future objectives.

Given the potential applications of both time crystals and quantum chaos, particularly in fields like quantum sensing \cite{cabot2023continuous,montenegro2023quantum,fiderer2018quantum,zhang2024harnessingquantumchaosspinboson} and non-equilibrium thermodynamics \cite{igor_thermodynamics}, this new quasi-periodic time-crystal phase may provide intriguing insights and advancements in these areas.
\bigskip

\noindent \textbf{Acknowledgments}---. The authors thank C. Bruder, B. Bu\v{c}a, A. Mercurio, and F. Ferrari for their discussion on the model and numerical methods. P.S. acknowledges support from the Alexander von Humboldt Foundation through a Humboldt research fellowship for postdoctoral researchers.

\appendix
\onecolumngrid

\section{Mean-field equations and fixed point analysis \label{mean-field}}

In this section, we discuss the mean-field dynamics of the system described by Eq.~(2) of the main text.
The classical system of dynamical equations for the expectation values of re-scaled spin operators $\hat{m}^{A,B}_{x,y,z}= \hat{S}^{A,B}_{x,y,z}/S $  can be obtained with the approximation $\langle \hat{m}^{\alpha}_{\mu} \hat{m}^{\beta}_{\nu} \rangle=\langle \hat{m}^{\alpha}_{\mu} \rangle  \langle \hat{m}^{\beta}_{\nu} \rangle$ since $[\hat{m}^{\alpha}_i,\hat{m}^{\alpha}_j] \rightarrow 0$ in the thermodynamic limit $S\rightarrow \infty$.
Therefore, corresponding set of mean-field equations for $m^{A,B}_{x,y,z}=\langle \hat{m}^{A,B}_{x,y,z} \rangle$ is given by
\begin{align}
    \frac{dm_x^{\alpha}}{dt} & = \kappa_\alpha m_x^{\alpha} m_z^{\alpha} + \Gamma m_z^{\alpha} m_y^{\beta\neq\alpha}, \nonumber\\
    \frac{dm_y^{\alpha}}{dt} & = -\Omega_{\alpha} m_z^{\alpha} + \kappa_\alpha m_y^{\alpha} m_z^{\alpha} - \Gamma m_z^{\alpha}  m_x^{\beta\neq\alpha}, \nonumber\\
    \frac{dm_z^{\alpha}}{dt} & = \Omega_{\alpha} m_y^{\alpha} - \kappa_\alpha \left[ m_x^{\alpha \, 2} + m_y^{\alpha \,2} \right] + \Gamma \left[ m_y^{\alpha}  m_x^{\beta\neq\alpha} - m_x^{\alpha}  m_y^{\beta\neq\alpha} \right], \label{eq:mfeqs}
\end{align}
where $\alpha,\beta \in \{A,B\}$ and $\kappa_A=-\kappa_B=\kappa$.

The dynamics of the system can be analyzed using fixed-point analysis. Fixed points are determined by solving the equations $\dot{m}^{A,B}_{x,y,z} = 0$. The behavior of the system near these fixed points is characterized by the eigenvalues of the Jacobian matrix $\mathcal{J}_{\mu,\nu}=d \mu / d \nu $ where $\mu,\nu \in \{m_{x,y,z}^{A,B}\}$. In the following, we discuss three fixed points of our system,  which describe the melted phase and different time crystal phases (CTC1, CTC2, and CTC3) shown in the phase diagram in Fig.~2 of the main text.

\begin{enumerate}
    \item First fixed point $\{ m^{*}_1 \}$: Let us first discuss the dynamics of the system around the following fixed point, 
    \begin{align}
         m^{A,B}_{x}=\pm \frac{\Gamma \Omega}{\Gamma^2-\kappa^2},~~~ m^{A}_{y}=-m^{B}_{y}=\pm \frac{\kappa \Omega}{\Gamma^2-\kappa^2},~~~ m^{A,B}_z=\pm \sqrt{1-\frac{\Omega^2(\Gamma^2+\kappa^2)}{(\Gamma^2-\kappa^2)^2}}.
    \end{align}
    
    A physical fixed point must follow $\vert m_\alpha^\mu \vert\leq 1$, and $  m_\alpha^\mu$ need to be real $\forall~ \mu ,\alpha$. This leads to the following constraints on the parameter values: $\vert \Gamma \Omega \vert \leq \vert \Gamma^2-\kappa^2\vert $, $\vert \kappa \Omega \vert \leq \vert \Gamma^2-\kappa^2\vert $ and $(\Gamma^2-\kappa^2)^2\geq \Omega^2 (\Gamma^2+\kappa^2)$. The last constraint is the lowest bound, which needs to be satisfied for the given point to be a physical fixed point. It can be further simplified as $\Omega\leq \vert \Gamma^2-\kappa^2 \vert/\sqrt{\Gamma^2+\kappa^2}$.  Therefore the given fixed point is only valid for $\Omega\leq \vert \Gamma^2-\kappa^2 \vert/\sqrt{\Gamma^2+\kappa^2}$.

    To understand the dynamics of the system in this parameter regime, we investigate the eigenvalues of the Jacobian matrix $\mathcal{J}$ which are given below
    \begin{equation}
        \Lambda_1=\{ 0,~~~0, ~~~\pm i\sqrt{\Gamma^2-\kappa^2}\sqrt{1-\frac{\Omega^2 (\Gamma^2+\kappa^2)}{(\Gamma^2-\kappa^2)^2}} \}.
    \end{equation}
 The eigenvalues are purely real for $\Gamma/\kappa<1$, which corresponds to the melted phase. The non-zero eigenvalues are purely imaginary for $\Gamma/\kappa>1$ which describes the CTC2 time crystal phase. These purely imaginary eigenvalues match with the frequency of the oscillations of the mean-field observables $m_{x,y,z}^{A,B}$ (see Fig.~3(h) of main text). 

    \item Second fixed point $\{ m^{*}_2 \}$: Another fixed point that exists for the system of coupled mean-field equations is given as follows
    \begin{align}
        m_x^A &=\frac{\Gamma(-\kappa Y+\Gamma \sqrt{X})}{\Omega\kappa Z},~~~ m_y^A=\frac{\kappa Z-\Gamma \sqrt{X}}{\Omega Z},~~~ m_z^A=-\frac{1}{\kappa}\sqrt{\frac{X+2\Gamma\kappa\sqrt{X}}{Z}},\nonumber \\
        m_x^B &= \frac{-\Gamma Y+\kappa \sqrt{X}}{\Omega Z},~~~~~ ~~m_y^B=\frac{\kappa Y - \Gamma \sqrt{X}}{\Omega Z},~~~m_z^B=0,
    \end{align}
    where $X=\Omega^2 Z-Y^2$, $Y=\Gamma^2-\kappa^2$ and $Z=\Gamma^2 + \kappa^2$. The $\{m_2^*\}$ is a physical fixed point only for the parameter regime $ \sqrt{\Gamma^2+\kappa^2}\geq \Omega \geq \vert \Gamma^2-\kappa^2 \vert/\sqrt{\Gamma^2+\kappa^2}$. 

    The corresponding eigenvalue of the Jacobian matrix is given by 
    \begin{equation}
        \Lambda_2=\{ 0,~~~0,~~~ \pm i\vert\Gamma^2-\kappa^2\vert \sqrt{\frac{\Omega^2 (\Gamma^2+\kappa^2)}{(\Gamma^2-\kappa^2)^2}-1},~~~- \sqrt{\frac{-X+2\Gamma\kappa\sqrt{X}}{Z}}\}.
    \end{equation}
    Here, the last eigenvalue is degenerate with degree two and is always real but negative. Thus, the dynamics of the system is decaying along the directions described by the corresponding eigenvectors of the $\mathcal{J}$. Meanwhile, other non-zero eigenvalues are purely imaginary for the given parameter range of the fixed point $\{m_2^*\}$. This fixed point corresponds to the CTC3 time crystal phase where the system initialized with any arbitrary state first spirals down to the given fixed point and then oscillates around it with the frequency given by the purely imaginary eigenvalues, as shown in Fig.~3(c,f,i).

    \item Third fixed point $\{m_3^*\}$: This is a trivial fixed point of the uncoupled time crystal phases and is given as follows
    \begin{align}
        m_x^{A,B}=\pm \sqrt{1-(\kappa/\Omega)^2},~~~m_y^A=-m_y^B=\kappa/\Omega,~~~m_z^{A,B}=0 .
    \end{align}

    The above fixed point is valid for $\Omega/\kappa>1$. Corresponding Jacobian eigenvalues are
    \begin{equation}
        \Lambda_3 = \{0,~~~0,~~~\pm \sqrt{\kappa^2-\Omega^2}\},
    \end{equation}
    where the non-zero eigenvalues are purely imaginary for the given parameter regime. This fixed point corresponds to the CTC1 time crystal phase and describes the exact dynamics for the uncoupled time crystals. The dynamics get modified in the presence of non-linear coupling, and the system exhibits quasi-periodic dynamics for $\Gamma\neq 0$, as depicted in Fig.~3(a,d) of the main text. The main peak of the Fourier transform of the dynamics obtained from the mean-field operators $m_z^\alpha$ matches with the purely imaginary eigenvalues of the given fixed point $m_3^*$, as shown in Fig.~3(g) of the main text.

\end{enumerate}

\begin{figure}[b!]
    \centering
    \includegraphics[width=0.9\textwidth]{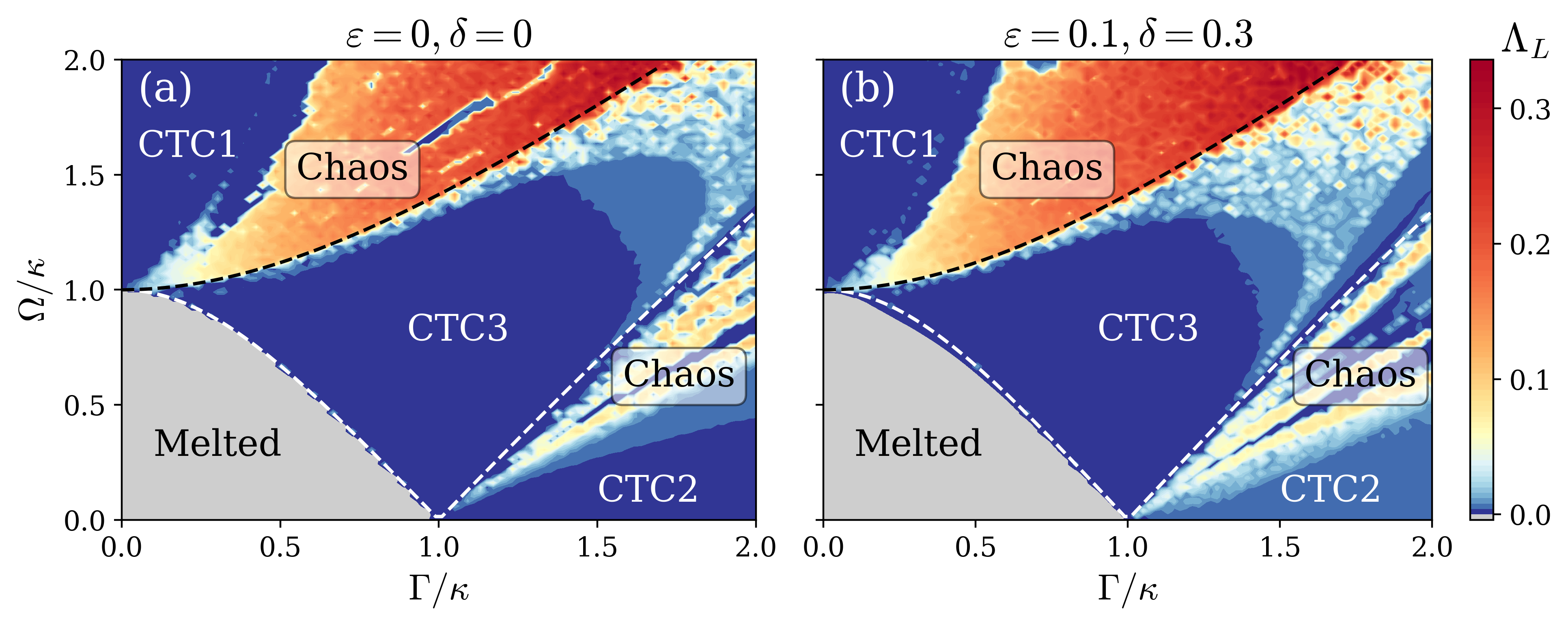}
    \caption{Maximum Lyapunov exponent for different initial states $m^{A,B}_x=\varepsilon,m^{A,B}_y=\delta,m^{A,B}_z=\sqrt{1-\varepsilon^2-\delta^2}$. Positive values of $\lambda_L$ indicate quasi-time crystal phase, $\lambda_L=0$ indicates time-crystal phases, and the system exhibits a time-independent melted phase for $\lambda_L<0$. All the phases are robust to different choices of initial states. The white and black dashed line corresponds to $\Omega/\kappa=\vert (\Gamma/\kappa)^2-1\vert/\sqrt{(\Gamma/\kappa)^2+1} $ and $\Omega/\kappa=\sqrt{(\Gamma/\kappa)^2+1}$ respectively, obtained by fixed point analysis.}
    \label{fig:chaos_lyapunov}
\end{figure}

\section{Lyapunov exponent and the phase diagram \label{lyapunov}}

Since fixed-point analysis relies on linearizing the system around its fixed points, it only provides insight into the local behavior near these points. 
As a result, the true dynamics of the full non-linear system can differ significantly, particularly in regions far from the fixed points or when non-linear effects dominate. 
Therefore, while fixed-point analysis is useful for understanding local stability, it may not capture the complete dynamics of the system.
To capture the global behavior of the system, we use Lyapunov exponents, which provide information about the overall stability and long-term dynamics of the system across its entire phase space.
In a quantitative sense, the separation between two nearby trajectories in phase space, starting with an initial difference $\delta \Vec{X}_0$ can be described as follows
\begin{equation}
    \vert \delta \Vec{X}(t) \vert \approx e^{\lambda t} \vert\delta \Vec{X}_0 \vert ,
\end{equation}
where $\lambda$ is the Lyapunov exponent.
The separation rate may vary depending on the orientation of the initial separation vector. 
As a result, there is a range of Lyapunov exponents corresponding to the dimensionality of the phase space. 
The largest of these Lyapunov exponent provides insight into the system's predictability.
The largest Lyapunov exponent $\lambda_{L}$ is always positive for a chaotic phase and describes the divergence rate of two nearby trajectories.
The $\lambda_{L}$ is zero for a limit cycle (time crystal phase) and negative for a stable fixed point.
We map out a phase diagram using the largest Lyapunov exponent, as shown in Fig.~\ref{fig:chaos_lyapunov} of the main text.

Here, we also discuss the effect of choosing different initial states on the phase diagram. 
In Fig.~\ref{fig:chaos_lyapunov}, we draw phase diagrams for two different initial states. 
The phase diagram is qualitatively similar for both initial states, and the phase boundaries predicted by the fixed point analysis remain unchanged. 
Only a quantitative change in the boundaries of chaotic/quasi-periodic phases can be observed in Fig.~\ref{fig:chaos_lyapunov}(a) and (b).
This confirms the robustness of the phase diagram with respect to the choice of different initial states.

\begin{figure}[htp!]
\centering
\includegraphics[width=\textwidth]{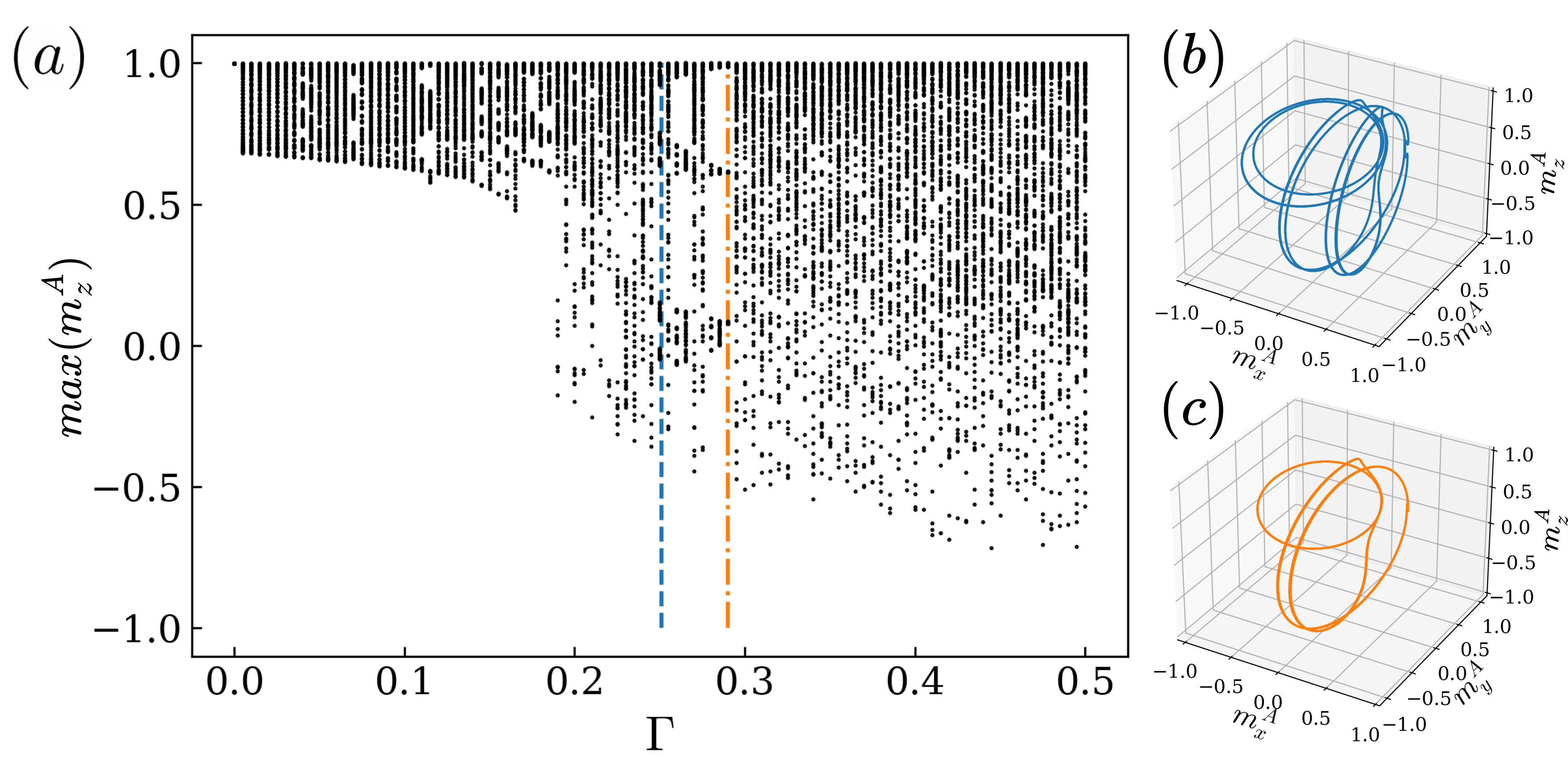}
\caption{(a) Period bifurcation diagram using the maximum values of $m_z^A$. Sub-figures (b) and (c) show that the period bifurcation leads to a period-6 and period-3 limit cycle behavior corresponding to $\Gamma/\kappa=0.251$ and $\Gamma/\kappa=0.29$ (represented by the number of maxima on the two vertical lines in (a)). Drive strength $\Omega/\kappa=1.2$ is the same for all sub-figures.}
\label{fig:bifurcation}
\end{figure}

A typical route to chaos is period bifurcation. 
In Fig.~\ref{fig:bifurcation}, we investigate the period bifurcation behavior by fixing the drive strength $\Omega/\kappa = 1.2$ and changing the coupling strength $\Gamma/\kappa$.
The maximum value of $m_z^A(t)$ shows that the system features a complex period-doubling bifurcation.
For $\Gamma/\kappa=0$, the system exhibits a single limit cycle phase, which becomes quasi-periodic for $\Gamma/\kappa\ll 1$.
Increasing the coupling strength $\Gamma/\kappa$ leads to a period-doubling behavior, and further increasing $\Gamma/\kappa$ eventually leads to chaotic dynamics for $\Gamma/\kappa>0.3$. 
Apart from the trivial limit cycle and chaotic phases, we could also observe several finite $n$-period limit cycles.
For example, Fig.~\ref{fig:bifurcation}(b) and Fig.~\ref{fig:bifurcation}(c) shows limit cycles with $n=6$ and $n=3$, respectively. 
Therefore, this coupled CTCs model features various interesting phases such as a simple limit cycle (time crystal phase), quasi-periodic dynamics, $n-$period limit cycles, and chaos.
\section{Truncated Wigner approximation \label{TWA-appendix}}

An approximate method to simulate a large open quantum spin system using truncated Wigner approximation (TWA) was introduced in \cite{Huber2021}. The truncated Wigner approximation for our model involves mapping the spin system to the bosonic system using the Schwinger transformation and applying the TWA on the bosonic system along with an additional positive diffusion approximation.
The Schwinger transformation is defined as the following
\begin{equation}
    S_+^i=a_i^\dagger b_i, ~~~S_-^i=a_i b_i^\dagger, ~~~S_z^i=(a_i^\dagger a_i-b_i^\dagger b_i)/2.
\end{equation}
This system of coupled bosonic operators follows the same $su(2)$ algebra as spin operators and the dynamics is dictated by:
\begin{equation}
    \dot\rho=-i\Big[\frac{\Omega}{2} (a_1^\dagger a_2+a_1 a_2^\dagger)+\frac{\Omega}{2} (b_1^\dagger b_2+b_1 b_2^\dagger)+\frac{\Gamma}{S}(a_1^\dagger a_2 b_1 b_2^\dagger+a_1 a_2^\dagger b_1^\dagger b_2),\rho\Big]+\frac{\kappa_A}{S}D[a_1 a_2^\dagger]+\frac{\kappa_B}{S}D[b_1^\dagger b_2].
\end{equation}
Now, we can easily apply the TWA to obtain the Fokker-Planck equation (FPE) for the Wigner function, which describes the dynamics of our system beyond the mean-field level for larger system sizes $S\gg 1$. 
The FPE of the Wigner function $W(\alpha_1,\alpha_2,\beta_1,\beta_2,t)$ can be defined as follows
\begin{equation}
    \frac{\partial W(\Vec{x},t)}{\partial t} = \Big[ -\frac{\partial}{\partial x_j}A_j(\Vec{x}) + \frac{1}{2} \frac{\partial}{\partial x_i} \frac{\partial}{\partial x^*_j} D_{ij}(\Vec{x})\Big]  W(\Vec{x},t)
\end{equation}
where $A$ is drift matrix and $D$ is diffusion matrix. The above FPE is obtained by the following mapping
\begin{align}
    a\rho  &= \Big[ \alpha + \frac{\partial}{2\partial \alpha^* } \Big] W(\alpha,t), \nonumber \\
    a^\dagger\rho  &= \Big[ \alpha^* - \frac{\partial}{2\partial \alpha } \Big] W(\alpha,t), \nonumber \\
    \rho a^\dagger  &= \Big[ \alpha^* +\frac{\partial}{2\partial \alpha } \Big] W(\alpha,t), \nonumber \\
    \rho a &= \Big[ \alpha - \frac{\partial}{2\partial \alpha^* } \Big] W(\alpha,t),
\end{align}
where the state of the system is now described by the phase space distribution function $W$, and the action of a bosonic operator is represented by a corresponding differential operator. The above mapping can be easily extended to the FPE for our model after applying the truncated Wigner approximation along with the positive diffusion approximation, which    is defined as follows
\begin{align}
    \frac{\partial }{\partial t}W(\alpha_2,\alpha_2,\beta_1,\beta_2,t) &= -i\Big[-\frac{\partial}{\partial \alpha_1}(\frac{\Omega}{2}+\frac{\Gamma}{S} \beta_1 \beta_2^*)\alpha_2  -\frac{\partial}{\partial \alpha_2}(\frac{\Omega}{2}+\frac{\Gamma}{S} \beta_1^* \beta_2)\alpha_1  - \frac{\partial}{\partial \beta_1}(\frac{\Omega}{2}+\frac{\Gamma}{S} \alpha_1   \alpha_2^*)\beta_2  \nonumber \\  &-\frac{\partial}{\partial \beta_2}(\frac{\Omega}{2}+\frac{\Gamma}{S} \alpha_1^* \alpha_2)  \beta_1 \Big]W+ \Big[\sum_{\mathcal{O}=\{\alpha,\beta\}}(\delta_{\mathcal{O}\alpha}-\delta_{\mathcal{O}\beta})\frac{\kappa_\mathcal{O}}{S} \Big(\frac{\partial}{\partial \mathcal{O}_1} (\vert \mathcal{O}_2\vert^2 + \frac{1}{2})\mathcal{O}_1 -\frac{\partial}{\partial \mathcal{O}_2} (\vert \mathcal{O}_1\vert^2 - \frac{1}{2})\mathcal{O}_2 \nonumber \\&+\frac{\partial}{\partial \mathcal{O}_1\partial\mathcal{O}_1^*} (\vert \mathcal{O}_2\vert^2 + \frac{1}{2}) +\frac{\partial}{\partial \mathcal{O}_2\partial\mathcal{O}_2^*} (\vert \mathcal{O}_1\vert^2 - \frac{1}{2}) \Big) \Big]W.
\end{align}

This FPE can be translated into the following equivalent collection of stochastic (Ito) differential equations
\begin{align}
    d \alpha_1 = -i(\frac{\Omega}{2}+\frac{\Gamma}{S} \beta_1 \beta_2^*)\alpha_2 dt - \frac{\kappa_A}{S}\Big(\vert \alpha_2 \vert^2+\frac{1}{2}\Big)\alpha_1 dt +\sqrt{\frac{\kappa_A}{2S}\Big(\vert \alpha_2 \vert^2+\frac{1}{2}\Big)}(d\mathcal{W}_1+id\mathcal{W}_2),\nonumber \\
    d \alpha_2 = -i(\frac{\Omega}{2}+\frac{\Gamma}{S} \beta_1^* \beta_2)\alpha_1 dt + \frac{\kappa_A}{S}\Big(\vert \alpha_1 \vert^2+\frac{1}{2}\Big)\alpha_2 dt +\sqrt{\frac{\kappa_A}{2S}\Big(\vert \alpha_1 \vert^2+\frac{1}{2}\Big)}(d\mathcal{W}_3+id\mathcal{W}_4),\nonumber \\
    d \beta_1 = -i(\frac{\Omega}{2}+\frac{\Gamma}{S} \alpha_1 \alpha_2^*)\beta_2 dt + \frac{\kappa_B}{S}\Big(\vert \beta_2 \vert^2+\frac{1}{2}\Big)\beta_1 dt +\sqrt{\frac{\kappa_B}{2S}\Big(\vert \beta_2 \vert^2+\frac{1}{2}\Big)}(d\mathcal{W}_5+id\mathcal{W}_6),\nonumber \\
    d \beta_2 = -i(\frac{\Omega}{2}+\frac{\Gamma}{S} \alpha_1^* \alpha_2)\beta_1 dt - \frac{\kappa_B}{S}\Big(\vert \beta_1 \vert^2+\frac{1}{2}\Big)\beta_2 dt +\sqrt{\frac{\kappa_B}{2S}\Big(\vert \beta_1 \vert^2+\frac{1}{2}\Big)}(d\mathcal{W}_7+id\mathcal{W}_8),
\end{align}
where $\mathcal{W}_i$ are independent Wiener processes with $\langle d\mathcal{W}_i d\mathcal{W}_j\rangle=\delta_{ij}dt$.
This equation appears in the main text.
\end{document}